%% file: main.tex
\documentclass[conference]{IEEEtran}
\IEEEoverridecommandlockouts
\usepackage{cite}
\usepackage{amsmath,amssymb,amsfonts}
\usepackage{algorithmic}
\usepackage{graphicx}
\usepackage{textcomp}
\usepackage{tabularx}
\usepackage{xcolor}
\usepackage{fontawesome}
\usepackage{pgfplots}
\usepackage[para]{footmisc}
\usepackage{dblfloatfix} 
\usepackage{tcolorbox}
\def\BibTeX{{\rm B\kern-.05em{\sc i\kern-.025em b}\kern-.08em
    T\kern-.1667em\lower.7ex\hbox{E}\kern-.125emX}}

 \newenvironment{quote1}{%
   \list{}{%
     \leftmargin%0.5cm   % this is the adjusting screw
        \rightmargin%0.5cm
   }
   \item\relax
}
{\endlist}

 \newenvironment{quote2}{%
   \list{}{%
     \leftmargin\parindent%0.5cm   % this is the adjusting screw
        \rightmargin\parindent%0.5cm
   }
   \item\relax
}
{\endlist}

\newcommand{\myquote}[2]{\begin{quote1}{\small\faComment}~\emph{#1}\end{quote1}} 
\newcommand{\rqquote}[1]{\begin{quote2}{#1}\end{quote2}} 
    
\begin{document}

\title{Under the Bridge: Trolling and the Challenges of Recruiting Software Developers for Empirical Research Studies}

\author{\IEEEauthorblockN{1\textsuperscript{st} Ella Kokinda}
\IEEEauthorblockA{\textit{School of Computing} \\
\textit{Clemson University}\\
Clemson, United States \\
ekokind@clemson.edu}
\and
\IEEEauthorblockN{2\textsuperscript{nd} Makayla Moster}
\IEEEauthorblockA{\textit{School of Computing} \\
\textit{Clemson University}\\
Clemson, United States \\
mmoster@clemson.edu}
\and
\IEEEauthorblockN{3\textsuperscript{rd} James Dominic}
\IEEEauthorblockA{\textit{School of Computing} \\
\textit{Clemson University}\\
Clemson, United States \\
domini44@clemson.edu}
\and
\IEEEauthorblockN{4\textsuperscript{th} Paige Rodeghero}
\IEEEauthorblockA{\textit{School of Computing} \\
\textit{Clemson University}\\
Clemson, United States \\
prodegh@clemson.edu}
}

\maketitle

\begin{abstract}
Much of software engineering research focuses on tools, algorithms, and optimization of software. Recently, we, as a community, have come to acknowledge that there is a gap in meta-research and addressing the human-factors in software engineering research. Through meta research, we aim to deepen our understanding of online participant recruitment and human-subjects software engineering research. In this paper we motivate the need to consider the unique challenges that human studies pose in software engineering research. We present several challenges faced by our research team in several distinct research studies, how they affected research, and motivate how, as researchers, we can address these challenges. We present results from a pilot study and categorize issues faced into three broad categories including participant recruitment, community engagement, and data poisoning. We further discuss how we can address these challenges and outline the benefits a full-study could provide to the software engineering research community. 

\end{abstract}


\input{sections/01_intro}
\input{sections/02_background}

\input{sections/03_method}
\input{sections/04_results}
\input{sections/05_discussion}

\section*{Acknowledgements}
Thank you to past and present members of our research group who took the time to help compile experiences from past studies and reflect on difficult studies. 

\bibliographystyle{IEEEtran}  
\bibliography{blopblip} 

\end{document}

%% file: sections/01_intro.tex
\section{Introduction}
As part of the software engineering research community, we need to research software developers and their behavior within the profession. For early career researchers and graduate students, recruiting software developers is an arduous task, with researchers turning to online participant recruitment. Online recruitment methods vary from cold-emailing or messaging, posting on social media like Facebook, Twitter, or LinkedIn, as well as using community social platforms like Slack, Discord, Gitter, and Reddit. However, new and veteran researchers alike may unexpectedly encounter toxic responses from social media users during recruitment, seemingly out of nowhere and without an avenue to rectify the situation they are now in. Toxicity is well studied in social media use and within leadership contexts \cite{almerekhi2020investigating, mohan2017impact, blackburn2014stfu, pelletier2010leader, carpenter2012narcissism, golbeck2017large}. However, it is rarely shown or spoken of in a research context. How do researchers handle their work being undermined before it is published? How do we rectify push-back from communities where target participants reside? How do we approach developers and communities without alienating or overstepping boundaries? Finally, how should the research community report these challenges?

Recent meta-studies make strides toward understanding and defining improved strategies for participant recruitment, however, many do not address encountering difficult participants and unexpected data \cite{rainer2022recruiting}. Our preliminary and proposed research aims to work hand-in-hand with recent studies regarding participant recruitment, help address the challenges that participant responses and behaviors pose, and provide transparency in research processes. Our goal is to invite the research community to provide feedback in the initial stages of this work. We seek to expand this work, find successful solutions to these challenges, and encourage those facing similar struggles to reach out. By understanding the challenges researchers face in online recruitment and data collection for human studies, we can inform and influence methods and approaches for participant recruitment and better understand empirical software engineering research approaches. 

% how successful are online recruitment methods?
% what recruitment methods do we use for SE studies

%% file: sections/02_background.tex
\section{Background}
In this section, we discuss the importance of meta research and its impacts on improving research methodology. Additionally, we discuss participant recruitment in social sciences. 

\subsection{Importance of Meta Research}
 Meta research ``aims to evaluate and improve research practices,'' and we believe that investigating our own research practices ensures effective and credible results. \cite{ioannidis2015meta}. Understanding appropriate research methods is foundational to study success, however, it has been often over-looked when applying across research disciplines \cite{ioannidis2015meta}. Through the analysis and adapting the methods of other research disciplines, such as social sciences and psychology, software engineering researchers can adapt research methodologies to their research studies. 

\subsection{Participant Recruitment in Software Engineering Research}

% Medical research has seen similar challenges of finding appropriate participants, including facing those who may be reluctant to participate \cite{streeton2004researching

% need something on appropriate particpants
%overfishing
%participant saturdation
% research on knowing when you have enough participants 

Historically, research shows convenience sampling to be the most predominant sampling strategy within software engineering research and that many participants come from a few companies in industry \cite{bouraffa2020two}. These historical sampling strategies raise concerns for generalizablity and and overall appropriateness of sample size. Recently, many researchers use paid platforms or other hosted sampling services and found that many of the same challenges sampling and recruiting appropriate and relevant participants and screen question challenges \cite{ebert2022recruiting, russo2022recruiting}. Additional recent work aims to shed light on addressing recruitment challenges surrounding sampling, participant screening, and using student software engineers as proxies for full-time software developers~\cite{endres2022making, danilova_you_2021, smith2022meta, brandt2022strategies}. Finally, it is important to note prominent one example of recruitment tension, and ongoing discussion on recruitment practices between researchers and software developers, where researchers were using a third-party site GHTorrent to solicit names and emails of developers \cite{gousios_deck2016,gousios2016}.

%% file: sections/03_method.tex
\begin{table*}[]
\centering
\caption{Breakdown tools and methods used to recruit participants for studies. \\{\scriptsize * - denotes in-person portion of the study, ** - denotes remote portion of the study}}
%    {\tabular[t]{@{}l@{}}* - denotes in-person portion of the study, ** - denotes remote portion of the study \endtabular}}
\begin{tabular}{c|l|l|l}
\hline
\multicolumn{1}{l|}{Study \#} & Ref.                     & Recruitment Details            & Challenges                                   \\ \hline
1                             & \cite{moster2022zone} & Reddit                         & Participant Recruitment, community engagement \\
2 & Under Review                   & Reddit, Discord, Other Social media & Participant Recruitment, Community Engagement, Data Poisoning \\
3                             & Under Revisions          & Email, Twitter Direct Messages & Community Engagement                          \\
4                             & Under Review             & Personal Contacts              & Participant Recruitment                      \\
5.1*                          & \cite{dominic2020remote} & Flyers, Social Media           & Participant Recruitment                      \\
5.2**                         & \cite{dominic2020remote} & Social Media                   & Participant Recruitment                      \\
6                             & Confidential Research      & Social Media, Flyers           & Data Poisoning                               \\
7 & Accepted, to be published 2023 & Reddit, Linked-In, Twitter          & Participant Recruitment, Community Engagement, Data Poisoning \\ \hline
\end{tabular}
\label{tab:studies}
\end{table*}

\section{Research Design}
In this section, we outline our research goals for both the pilot study and subsequent research, as well as, methods approach for a full study.

\subsection{Research Goals}
Motivated by our experiences recruiting software developers for research studies, we believe the following research questions (RQs) can establish a baseline approach for understanding the recruitment challenges and their effects on research. The goal of our meta-research within software engineering is to deepen our understanding of approaches to online developer-participant recruitment and bring awareness to challenges faced by software engineering researchers which may not be readily apparent. 

\rqquote{{$RQ_1$:} What challenges do software engineering researchers encounter when recruiting human participants for research studies?}
\rqquote{{$RQ_2$:} How have these challenges impacted research?}
\rqquote{{$RQ_3$:} How have these challenges impacted researchers' personal well-being?}
\rqquote{{$RQ_4$:} What lessons have software engineering researchers learned from recruitment challenges encountered?}

The rationale behind $RQ_1$ is to understand the extent and magnitude of challenges present in human-subjects software engineering research. From there, $RQ_2$ and $RQ_3$ aim to understand the impact of challenges on researchers. Finally, $RQ_4$ addresses how researchers can lessen the impact of challenges and understand what has helped when a challenge is posed during research.  

\subsection{Qualitative Research Approach}
\subsubsection{Pilot Study} For our pilot study, authors and previous members of our research group came together to share experiences and challenges encountered while recruiting for software development research. In total, we analyzed 7 different research studies and found shared and unique challenges from these studies.  From there, the team classified challenges into 3 broad categories: participant recruitment, community engagement, and data poisoning. Table \ref{tab:studies} provides a breakdown of the research studies and the challenges they pose.

\subsubsection{Proposed Study} Using the RQs from our pilot study, we propose a survey and interview-based approach to meta-research within software engineering researchers. Our proposed survey will elicit both quantitative and qualitative data points regarding challenges and quantify their impacts on research from software engineering researchers who recruit human participants. We propose using standard statistical analysis for quantitative questions using methods outlined in \cite{ali2016basic}. For qualitative data from the survey and interviews we propose using open-coding techniques \cite{corbin2014basics}.

%% file: sections/04_results.tex
\section{Preliminary Results}
In this section we outline experiences and challenges faced by the authors while recruiting human subjects in software engineering studies. Additionally we highlight the impacts of the challenges on research and well-being.

\subsection{Challenges}
Several types of challenges emerged from authors' experiences with many challenges surrounding approaching communities to solicit research, response from communities, concerns over oversampling from niche participant pools, and participants purposefully subverting their data. 

\vspace{0.1cm}\noindent\textbf{\emph{Participant Recruitment --}} Necessary to human-subjects research are the participants, and as researchers we run into several issues with recruitment, especially online recruitment. One of the largest problems faced by our researchers, across all studies for in-person events or interviews were no-shows or ghosting\footnote{Ghosting is ending contact with someone after prior communication or agreement without justification and any subsequent followup or acknowledgement}. Studies S3, S4, S5.1, and S7 faced the challenges of participants not showing up or ghosting after reaching out to reschedule. Participants who do not show up to an agreed time can hinder research progress such as study cancellations, demotivate future research projects, and can put researchers in negative or awkward situations with participants who arrived on time. In Study 5.1, in-person research needed to be cancelled due to lack of participants, when emailing with a weeks notice to those who signed up, researchers faced backlash from a participant that emailed over the course of several hours:

%lead to study cancellations that hinder research progress and be demotivating and lead to awkward situations with those who did sign up. 
\myquote{``You are doing research for A university that regularly researches stuff. You’re telling me you have no money. Riiiighhhtttttt. Go f*** yourself.''}

\myquote{``I will personally make sure that everyone knows not to waste their time on [University]’s little programming thing that you have. My Venmo is @[VenmoHandle] if you want to compensate me.''}

\myquote{``I actually am a [Rival University] fan and this is the most Trump thing I could of possibly expected for from a good Ole Boys.''}

\myquote{``Just don’t try to bring your bulls*** surveys to [city]. We have a beach and don’t need to be dealing with snowflakes.''}

Additionally, within S5.1 and S5.2, recruiting paid participants with appropriate experience for a software development study proved challenging. Participants expressed interest in the study but lacked the appropriate programming experience to participate, with 9 out of 89 initial participants (10\%) ultimately being rejected from the study. With niche participant requirements or unique research interests, each researcher encountered concerns surrounding over-fishing problems (reaching out to the same communities for participants) and ultimately reaching enough participants to make research generalizable. 

% recruiting appropriate experienced participatns
% over-fishing
% ghosting
% rude responses when participants dont get their way

\vspace{0.1cm}\noindent\textbf{\emph{Community Engagement --}} In the case of S1, S2, and S6 recruitment using online communities posed several issues regarding community rules related to research solicitations, community moderator responses, and online platform automated spam filters. When reaching out to online communities it is best to understand the community norms and rules, which some define clearly and others do not. We found that many communities who would be appropriate to invite to participate in research do not allow for these types or solicitations or to contact the moderation team for permission. When contacting the moderator teams for permission when communities ask or when rules are unclear, we have faced several challenges - no responses and hostile reactions. In S7, the research team found that 202 out of 1000 (20.2\%) subreddit moderator teams reached out to were willing to allow advertisements or post. When presenting to moderators our team would give an overview of the research and its affiliated university, why we are reaching out, outline benefits and protections for participants, state its IRB approval, and often a link to the survey for it to be previewed. While most communities would not respond or politely decline, others took a hostile approach with verbal abuse and name calling:

\myquote{``You are dogsh**.''}

\myquote{``You are thieves stealing from this website, thats a fact''}

\myquote{``Come back with cooler ideas and maybe I will reconsider.''}

\myquote{``No, and [if] you post I will remove it.''}

Additionally, we found we would be banned from communities for reaching out to moderation inboxes with no further explanation or communication from the moderation team.

% Finding ways to reach out to the community moderators for approval to post
% Rude responses from community moderators
% getting caught in automated spam filters

\vspace{0.1cm}\noindent\textbf{\emph{Data Poisoning --}} While not a problem for everyone, several authors run into ``troll" or disruptive behaviors from research participants through their answers on survey questions and in responses to interview questions. In S7, researchers encountered those using gender self-description boxes to voice their opinions like \emph{``there are only two genders''} or non-sensible answers like \emph{``dinosaur.''} Another researcher, in S2, encountered hostile survey responses from consenting participants and aimed at subverting the research, with participant responses to questions such as \emph{``f*** you''} and \emph{``My answers have derailed this survey to the point my answers are probably meaningless to your research.''} 

% speaking to or addressing the researcher in answers
% 

\subsection{Impact of Challenges}
%what to do when you get troll answers for things like demographics -- do you throw out the whole response even if the rest seems genuine?
% uncertain if online participant recruit is worth it based on past experiences
Impacts from these challenges varied from researcher to researcher, but all expressed frustration and demotivating about the research. When data subversion occurred, in the cases of S7 and S2, researchers indicated uncertainty if subsequent data in those responses were genuine and should be included. Particularly in the case of S2, researchers reported that data from these participants could be used in some capacity but questioned overall inclusion of these responses, as they did not have a large impact on overall generalizability of results, and were often outliers.

For those who faced participant recruitment challenges, over-fishing the same communities, finding appropriate participants, and concerns over small sample size led researchers to question if online recruitment methods are worth the hassle and appropriate channels to pursue for research. Additionally, the frustrations of challenges led to overall demotivation to continue research, hesitation toward using certain social media platforms, and takes a toll on personal well-being.

%% file: sections/05_discussion.tex
\section{Discussion}

% impacts of difficult research - hard on new researchers, discouraging, feel that it is not talked about in the reserach community, transparency needed
Research emphasizes the need to reach broader and diverse participants and achieve larger representative sample sizes, researchers must understand how to filter, address, and/or use data from participants who may provide difficult data \cite{rainer2022recruiting}. From our example studies, participant response data that proved challenging, we did not want to completely exclude, but often did not make any point other than being outlier data. 

%talk about how we think we could improve when reaching out to communities
It is imperative to acknowledge that researchers may be the root cause of their own problems, and that hostile behavior from participants or communities may reside in how we, as researchers, interact, interface, and subsequently treat participants and communities. When soliciting developer communities to participate in research, we recommend to provide concise and digestible information about the study as well as clearly define why developers should participate when reaching out to communities. We found it helpful and respectful to reach out to moderation teams when questioning if research solicitations were appropriate for the community or when reaching out to individuals. Below we provide exemplar texts from reaching out to community moderators and individuals: 

\textbf{\textit{Community Moderators Boilerplate}}
\begin{quote}
    ``Hey [Community] Mods!
    \\ \\
    I am a human factors and software development researcher from [University] and am looking to recruit participants for an IRB approved research study on [details here].
    \\ \\
    This study focuses on [Additional detail of your study and why you are reaching out to this community]. From this research I am looking to understand [explain the benefits of your research and why participants from here should participate]
    \\ \\
    I am more than happy to answer any questions y’all may have, and I appreciate your time!
    \\ \\
    All data is anonymous. You can find the survey here: [link to survey].''
\end{quote}

\vspace{.2cm}

\textbf{\textit{Individual Boilerplate}}
\begin{quote}
    ``Hi [Individual's Name]!
    \\ \\
    My name is [Name Here], and I am a [Current Position] at [Affiliated University/Company] under advisement of [Advisor/Research Group] ([Links to personnel or research group homepages]). I am looking to recruit participants for an [Interview/Survey] that will take [Time Commitment] for research related to [topic]. I am really interested in [explain why you are reaching out to this person in particular]! We are investigating [things you think will help you move forward with your research with this individuals help]. 
    \\ \\
    What this interview would entail would be a short Q\&A session regarding [overview of what you plan to ask about]. 
    \\ \\ 
    I appreciate your time in reading this, and am more than happy to answer any questions you may have!''
    \\
\end{quote}

\begin{tcolorbox}[width=\linewidth, halign=center, colframe=black, colback=gray!5, boxsep=1mm, arc=3mm]
\noindent \centering\textbf{Recommendations based upon observations for Software Developer Research Recruitment}
\begin{itemize}
\item Introduce the work concisely and in plain, understandable language
\item Identify researchers, affiliated institutions
\item Explain why you are approaching this person or community
\item Motivate benefits of participation
\item Clear and concise impacts to participants, if any
\item Identify how you handle or anonymize data
\item Provide contact information for questions and concerns
\item Provide a link to survey or interview questions for moderator preview (if applicable)
\end{itemize}
\end{tcolorbox}
\vspace{.2cm}

These outlined recommendations for participant recruitment were used in part for studies 2, 3, and 7, and fully resulted from feedback we received from participants or community moderators while soliciting participants for these same studies.

Additionally, we believe that efforts to increase sample sizes and pull from diverse participant pools may help alleviate challenges from reaching out to communities, find broader participation and support from online software development communities, and lead to more generalizable research results.

Brown's research outlines a standalone website where researchers may post research opportunities along with unique incentives for participation \cite{brown2022nudging}. Additionally, we propose as part of our future work to extend presence on current social media outlets and create a community for software developers to engage and interact with researchers on proposed and upcoming work. Through this we hope to foster meaningful interactions between developers and researchers and address identified challenges in current software engineering research recruitment methods.

Though everyone faces challenges differently, we believe that providing transparency in research challenges will encourage new researchers, re-motivate those who feel discouraged, and compliment recent work with respect to participant sampling approaches. By exploring alternative methods to participant recruitment, we may also alleviate and lessen the impacts from these observed challenges. 

\section{\textcolor{black}{Future Plans}}

%address that services and commmuniteis like Reddit can be used to recruit particpants, cite Chris's portal

\textcolor{black}{For future work, we propose a two-part study using interviews and surveys continuing meta-analysis within the software engineering research community investigating research methods and researcher approaches to communities, as well as working with software developer community moderators and leaders to understand how researchers approach and interface with their communities. Through this we hope to understand how and why hostile responses from participants and communities may arise, and what, if anything, researchers can do to mend relationships with participants and communities if given the opportunity.}

\textcolor{black}{Continuing meta-analysis should investigate participation fatigue - how often are researchers reaching out to communities, do communities feel spammed or exhausted in being continually asked for their time? It may be beneficial to work with these communities to understand the types of requests they are receiving from researchers and non-researchers alike who are turning to these communities for input. Understanding what is asked of these communities with respect to their participation in research can help researchers and these communities form a trustworthy, and mutually beneficial relationship. However, as researchers, we need to understand that some communities may not work to work with us, they do not owe us their participation or time, and we must respect that.}

\textcolor{black}{Future work will self-reflect and compare to other research disciplines on our approaches to participant recruitment and relationships with developer communities. Are participant recruitment methods clear, concise, and relevant to the developers we want to reach? Are recruitment solicitations upfront about their impact on participant developers and justify time spent participating?}

%\textcolor{black}{Finally, we propose and are working on creating an online community where developers and software engineering researchers may co-mingle and share...}

%Within this proposed research, we must respect personal and sensitive issues that may arise during data collection and anonymous and disambiguate data from any one participant or community.

\section{Conclusion}
By sharing our research teams' experiences, we highlight the challenges that software engineering researchers face and show how these challenges impact research. We believe that by sharing these experiences, the software development research community will come together and share their experiences with online recruitment and human subjects research, and contribute to progressing research methodologies and strategies for future research. Through subsequent research, our work will contribute to re-framing recruitment methods and engage in research practices that are more generalizable and reflective of the populations we study. Finally, we invite feedback and collaboration from researchers in the software engineering community who may have experienced similar recruitment difficulties and want to collaborate towards meaningful contributions and solutions.

% we invite others to share their experiences and look into cross-referencing methods of human particpant reserach within software engineering resarch spheres